\documentclass[twocolumn,showpacs,preprintnumbers,amsmath,amssymb]{revtex4}

\usepackage{graphicx}% Include figure files

\begin{document}

\title{Time Dependent Pairing Equations for 
Seniority One Nuclear Systems}

\author{M. MIREA}
\affiliation{Horia Hulubei National Institute for Physics and Nuclear Engineering, 
P.O. Box MG-6, 077125 Bucharest-Magurele, Romania}

\begin{abstract}
 When the time dependent 
Hartree-Fock-Bogoliubov intrinsic equations of motion are solved 
in the case of seniority one nuclear systems,
the unpaired nucleon 
remains on the same orbital. The blocking effect hinders the possibility to 
skip from one orbital to another. This unpleasant feature is by-passed
with
a new set of pairing time dependent equations that allows 
the possibility that the unpaired nucleon changes its single-particle level.
These equations generalize the time dependent Hartree-Fock-Bogoliubov equations
of motion by including the Landau-Zener effect. The derivation of these
new equations is presented in details.
These equations are applied in the case of a superasymmetric fission process, that is,
in order to explain the fine structure the $^{14}$C emission from $^{233}$Ra. 
A new version of the Woods-Saxon model extended for two-center
potentials is used in this context.
\date{today}
\end{abstract}

\maketitle

\section{Introduction} 

 In the Hartree-Fock (HF) approximation, the self-consistent
potential for heavy nuclei is quite smooth, since it includes the 
convolution of many
instantaneous densities. If the potential varies in time,
each single-particle wave function moves
independently in the smoothly varying well. The Pauli principle
is fulfilled being mediated permanently through the mean field potential. 
The two-body collisions are incorporated in the
equations of motion only to the extend to which they contribute
to the mean field. In principle, the time dependent HF approach
treats the residual interactions exactly only if the mean
field is allowed to break all symmetries. A such approach leads
to a very big computational problem. 
In order to avoid such numerical difficulties, usually, 
the HF mean field is constrained
to be at least axial symmetric. In this case, levels characterized by
the same good quantum numbers cannot intersect. When the equations 
of motion are solved in such circumstances, an individual single particle
wave function will belong to only one orbital characterized by some
good quantum numbers and the mechanism
of level slippage \cite{dlh} is not allowed. 
In general, two dynamical approaches are used. On one hand,
the generator coordinate
method assumes that the internal structure of the decaying system
is equilibrated at each step of the collective motion \cite{goutte}. On the
second hand, the exchange between collective and internal
degrees of freedom is neglected \cite{berger}, so that adiabaticity is assumed.
Usually, these behavior leads
to the unpleasant feature that a system, even moving infinitely slowly,
could not end up in its ground state. Some attempts were done to extend
the time-dependent HF method in order
to includes collision terms \cite{tang,grange}.
This
problem was partially solved by introducing a residual
pairing interaction
in the Time Dependent Hartree-Fock-Bogoliubov 
(TDHFB) approach \cite{schutte,koonin}.
This last method provides the possibility of level slippage 
for pairs and allows a
description of the nuclear dynamics. For example, in the case of an even-even
system, this method represents a tool to estimate the dissipation during
disintegration processes \cite{koonin,mirea1}. A connection
with the Landau-Zener effect is included in the TDHFB equations.
Pairs undergo Landau-Zener transitions on virtual levels
with coupling strengths given by the gap $\Delta$ \cite{blocki}.
Unfortunately, for seniority one nuclear systems the pairing
residual interaction does not affect a single nucleon and,
during the deformation of the nucleus from its initial state up to scission,
the unpaired particle remains located on the same orbital. The level
slippage is again forbidden for the blocked level.

In the case of independent single-particles, neglecting residual interactions,
the problem of the unpaired nucleon is solved 
in terms
of the Landau-Zener effect.  
The Landau-Zener effect reflects a mechanism that allows the possibility that 
a single nucleon skips 
from one single-particle level to another one
in some avoided level crossing regions. The 
probabilities that the unpaired nucleon arrives in different final states can 
be computed by solving a system of coupled channel equations that characterizes 
the microscopic motion. In the following, a way to introduce a 
similar mechanism for the unpaired 
nucleon in superfluid systems is investigated. 
The TDHFB equations will be 
generalized in order to include the Landau-Zener effect. The classical TDHFB 
equations and the equations that govern the Landau-Zener effect will be obtained 
as particular cases of the new time dependent pairing equations. 

\section{ Landau-Zener effect} 

The single-particle levels are function of the deformation parameters that 
characterize the shape of a nucleus. Levels characterized by the same quantum 
numbers associated to some symmetry of the system cannot cross and exhibit 
avoided level crossings. The transition probability of a nucleon from one 
adiabatic level to another one is strongly enhanced in an avoided 
crossing region.  This promotion mechanism is known as the Landau-Zener 
effect \cite{dlh}.

\begin{figure}
\resizebox{0.50\textwidth}{!}{
  \includegraphics{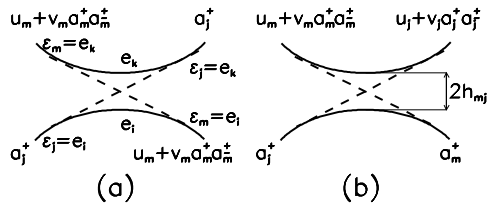}}
\caption
{
 Ideal avoided crossing region.
Possible transition states in an avoided crossing region for an 
unpaired nucleon in the superfluid model.
}
\label{figura2}
\end{figure}

In Fig. \ref{figura2}(a)  an ideal avoided crossing $(j,m)$
between two adiabatic levels $e_{k}$ and
$e_{i}$  
is displayed. The diabatic levels are $\epsilon_m$ and $\epsilon_j$. 
If the variation of 
the generalized coordinate is produced slowly and the nucleon is initially 
located on the level $e_i$, after the passage of the avoided crossing region, 
the nucleon will practically remain on the same adiabatic level. In this case, 
the motion is adiabatic and the nucleon follows the adiabatic state $e_i$. If the 
variation of the generalized coordinate is produced suddenly, then the nucleon 
will skip with a great probability on the adiabatic level $e_k$. The motion is diabatic, and 
the nucleon follows the diabatic state $\epsilon_j$. A formalism can be used to obtain 
the promotion probability. Assuming an $n$-state approximation, the wave function 
of the unpaired nucleon can be formally expanded in a basis of $n$ diabatic wave 
functions $\varphi(r)$ as follows:
\begin{equation}
\Psi(r,t)=\sum_{i}^{n}c_{i}(t)\varphi(r)\exp\left(-{i\over\hbar}
\int_{0}^{t}\epsilon_{i}(\tau)d\tau\right),
\label{dezv}
\end{equation}
where the matrix elements with the diabatic states are 
$\epsilon_{i}=<\varphi_{i}\mid H\mid \varphi_{i}>$,
$h_{ij}=<\varphi_{i}\mid H\mid \varphi_{j}>=h_{ji}$,
$H$ is the mean field Hamiltonian and $c_{i}$ are amplitudes. Inserting  
the wave function (\ref{dezv}) in the 
time-dependent Schr\"{o}dinger equation, 
the following system of coupled equations is obtained:
\begin{equation}
\dot{c}_{i}(t)={1\over i\hbar}\sum_{j\ne i}^{n}c_{j}(t)h_{ij}(t)
\exp\left(-{i\over\hbar}\int_{0}^{t}(\epsilon_{j}(\tau)-\epsilon_{i}(\tau))d\tau\right)
\label{lzu}.
\end{equation}
Here, $p_{i}=|c_i|^2$ is the probability to find the unpaired nucleon on the level $i$ .

These equations were already used to explain the resonant like structure
of the inelastic cross sections in heavy-ion collisions \cite{moon,park}, 
the fine structure
in cluster emission \cite{mirea14c,mireaac} and in alpha decay \cite{mireaalpha}.
the resonant structure in the fission cross section \cite{fis1,fis2,fis3}.

\section{ Superfluid systems}

An effect analogous to the Landau-Zener one can be obtained by generalizing
the TDHFB equations for the case of seniority one nuclear systems. 
The problem will be explored in the simplest
possible way: a monopole pairing force, and 
a sufficiently weak pairing such that the nucleons
are not redistributed to change significantly the mean field potential. 
In order to make the problem tractable, two approaches are investigated.
The first one is valid for a low lying levels system with
 a small number of avoided crossing
regions, so that variations of densities $\rho_{i}$ and pairing
moment components $\kappa_{i}$ due to the blocked
level can be neglected. The second one takes into account the blocking
effect, that is, the fact that
$\rho_{i(m)}$ and $\kappa_{i(m)}$ depend on the blocked level $m$. 

\subsection {Low lying levels}

Using quasiparticle creation and annihilation operators $\alpha_{\bar{k}}^{+}$ and 
$\alpha_{\bar{k}}$
\begin{equation}
\begin{array}{c}
\alpha_{k}=u_{k}a_{k}-v_{k}a_{\bar k}^{+};\\
\alpha_{\bar{k}}=u_{k}a_{\bar k}+v_{k}a_{ k}^{+};\\
\alpha_{k}^{+}=u_{k}a_{k}^{+}-v_{k}^{*}a_{\bar k};\\
\alpha_{\bar{k}}^{+}=u_{k}a_{\bar k}^{+}+v_{k}^{*}a_{ k};\\
\label{anho}
\end{array}
\end{equation}
it is possible to construct some interactions that help us to promote the nucleon 
from one diabatic level to another. The two situations plotted in Fig. \ref{figura2}
can be modeled. In the plot \ref{figura2}(a), the single particle
follows the diabatic level $\epsilon_{j}$ while in \ref{figura2}(b) it stays on 
the adiabatic one $e_{i}$. Here $a_{k}^{+}$ and $a_{k}$ denote operators for creating 
and destroying a particle in the state $k$, respectively. The state characterized 
by a bar signifies the time-reversed partner of a pair. The parameters 
$v_{k}$ and $u_{k}$ are the occupation and vacancy amplitudes, respectively. 
Because only the relative phase between the parameters $u_{k}$ and $v_{k}$ 
matters, in the following $u_{k}$ is considered as a real quantity and 
$v_{k}$ a complex one.
The interaction able to promote the unpaired nucleon from one adiabatic level to
another must be given by product of operators of the type (\ref{anho}).

In order to obtain the equations of motion, we shall start from the variational 
principle taking the Lagrangian as
\begin{equation}
\delta L=\delta <\varphi \mid H-i\hbar{\partial\over \partial t}
+H'-\lambda N\mid
 \varphi >,
\label{lagrange1}
\end{equation}
and assuming the many-body state formally expanded as a superposition of  
$n$ time dependent BCS seniority one diabatic wave functions
\begin{equation}
\varphi(t)=\sum_{m}^{n}c_{m}(t)a_{m}^{+}\prod_{l\ne m}\left(u_{l}(t)+
v_{l}(t)a_{l}^{+}a_{\bar{l}}^{+
}\right).
\label{functii1}
\end{equation}
The Lagrangian contains several terms. The first one is the many body 
Hamiltonian with pairing residual interactions 
\begin{equation}
H(t)=\sum_{k>0} \epsilon_{k}(t)(a_{k}^{+}a_{k}+a_{\bar k}^{+}a_{\bar k})-G\sum_{k,l>0}a_{k}^{+}a_{\bar{k}}^{+}
a_{l}a_{\bar{l}}.
\label{ham1}
\end{equation}
The residual interactions between diabatic levels characterized by the 
same quantum numbers that are responsible for the Landau-Zener effect are
assumed on the form:
\begin{equation}
\begin{array}{c}
H'(t)=\sum_{i,j\ne i}^{n}h_{ij}(t)
\alpha_{i}^{+}\alpha_{j}\\
=\sum_{i,j\ne i}^{n}h_{ij}(t)
(u_{i}a_{i}^{+}-v_{i}^{*}a_{\bar{i}})(u_{j}a_{j}-v_{j}a_{\bar{j}}^{+})
.
\end{array}
\label{hprim1}
\end{equation}
The sum runs over  
diabatic levels $i$ and $j$. The number of particle 
operator is:
\begin{equation}
N=\sum_{k>0}(a_{k}^{+}a_{k}+a_{\bar k}^{+}a_{\bar k})
\end{equation}
After some calculations, as detailed in Appendix \ref{appendixa}, the next time dependent 
coupled channel equations are obtained \cite{mireamod}:

\begin{equation}
i\hbar \dot{\rho}_{l}=\sum_{m}^{n}p_{m}
\left\{\kappa_{l}\Delta_{m}^{*}-\kappa_{l}^{*}\Delta_{m}\right\},
\label{tbcs1}
\end{equation}

\begin{equation}
i\hbar\dot\kappa_{l}=\sum_{m}^{n}p_{m}\left\{\left(2\rho_{l}-1\right)\Delta_{m}+
2\kappa_{l}\left(\epsilon_{l}-\lambda\right)\right\},
\label{tbcs2}
\end{equation}

\begin{equation}
i\hbar\dot{p}_{m}=\sum_{j\ne m}^{n} h_{mj}\left(S_{mj}-S_{jm}\right),
\label{tbcs3}
\end{equation}

\begin{equation}
\begin{array}{c}
i\hbar\dot{S}_{jm}=S_{jm}\left\{-{1\over G}\left(
\mid\Delta_{m}\mid^{2}-\mid\Delta_{j}\mid^{2}\right)+
\left(\epsilon_{m}(t)-\epsilon_{j}(t)\right)\right.\\
-{1\over 2}\left(-{\rho_{m}\over\kappa_{m}}+2\kappa_{m}^{*}+{\rho_{j}\over\kappa_{j}}
-2\kappa_{j}^{*}\right)\sum_{l}^{n}p_{l}\Delta_{l}\\
\left.-{1\over 2}\left(-{\rho_{m}\over\kappa_{m}^{*}}+2\kappa_{m}+{\rho_{j}\over\kappa_{j}^{*}}
-2\kappa_{j}\right)\sum_{l}^{n}p_{l}\Delta_{l}^{*}\right\}\\
+\sum_{l\ne m,j}^{n}\left[h_{ml}(t)S_{jl}-h_{jl}(t)S_{lm}\right]+
h_{mj}(t)(p_{j}-p_{m}).
\end{array}
\label{tbcs4}
\end{equation}

The next notations are used:
\begin{equation}
\begin{array}{c}
\Delta_{m}=G\sum_{k\ne m}\kappa_{k};\\
\Delta_{m}^{*}=G\sum_{k\ne m}\kappa_{k}^{*};\\
\kappa_{k}=u_{k}v_{k};\\
\rho_{k}=\mid v_{k}\mid^{2}\\
p_{m}=\mid c_{m}\mid^{2}\\
S_{jm}=c_{j}^{*}c_{m}.
\end{array}
\label{notatii}
\end{equation}
$\rho$ are single particle densities, $\kappa$ are pairing moment components, and $p_{m}$
denotes the probability to have an unpaired nucleon on the level $m$.
$\rho$ and $p$ are real quantities while $\kappa$ and $S$ are complex ones.
In analogy with $\kappa$, $S$ (having the property $\mid S_{jm}\mid^{2}=p_{j}p_{m}$)
 can be called as unpairing moment component.
Whenever the upper limit $n$ is specified for a sum, it is implicitly assumed that the
operation is realized on the $n$ possible diabatic states of the unpaired nucleon.
In this paper, the sum over pairs energy generally runs within
the index $k$. When the single-particle sum over $k$ is realized only for one partner
of each reversed pair the result is multiplied with a factor 2.
The index $k$ runs over a workspace that allows the pairing force to operate 
only over a a finite number 
of active levels around the Fermi energy.

\subsection{Blocking effect}

If the blocking effect is taken into consideration, each seniority
one BCS wave function is characterized by its own set of $\rho$ and $\kappa$ values,
and the trial wave function is:
\begin{equation}
\varphi(t)=\sum_{m}^{n}c_{m}(t)a_{m}^{+}\prod_{l\ne m}
\left(u_{l(m)}(t)+v_{l(m)}(t)a_{l}^{+}a_{\bar{l}}^{+}\right).
\label{wf2}
\end{equation}
The Landau-Zener interaction is postulated as follows:
\begin{equation}
\begin{array}{c}
H'(t)=\sum_{i,j\ne i}^{n}h_{ij}(t)\\
\times
\alpha_{i(j)}^{+}\alpha_{j(i)}
\prod_{k\ne i,j}\alpha_{k(j)}a_{k}^{+}a_{k}\alpha_{k(i)}^{+}\\
\\
=\sum_{i,j\ne j}^{n}h_{ij}(t)
(u_{i(j)}a_{i}^{+}-v_{i(j)}^{*}a_{\bar{i}})(u_{j(i)}a_{j}+v_{j(i)}a_{\bar{j}}^{+})\\
\times
\prod_{k\ne i,j}\alpha_{k(j)}a_{k}^{+}a_{k}\alpha_{k(i)}^{+}
,
\end{array}
\label{cort}
\end{equation}
where the quasiparticle creation and annihilation operators
\begin{equation}
\begin{array}{c}
\alpha_{k(j)}=u_{k(j)}a_{k}-v_{k(j)}a_{\bar k}^{+};\\
\alpha_{\bar{k}(j)}=u_{k(j)}a_{\bar k}+v_{k(j)}a_{ k}^{+};\\
\alpha_{k(j)}^{+}=u_{k(j)}a_{k}^{+}-v_{k(j)}^{*}a_{\bar k};\\
\alpha_{\bar{k}(j)}^{+}=u_{k(j)}a_{\bar k}^{+}+v_{k(j)}^{*}a_{ k};\\
\label{anho2}
\end{array}
\end{equation}
are now associated to each blocked level ($j$). This 
interaction considers only an approximate way to describe the full phenomenon. 
In this context, if a diabatic wave function
$i$ is "reflected" in an avoided crossing
region $(i,j)$, this wave function is transformed in a component of
the diabatic wave function $j$. The reality
is more complicated. When a diabatic wave function $i$ is "reflected" in an
avoided crossing region, this wave function must be split into two parts: a
transmitted diabatic wave function $i$ and a reflected adiabatic wave function $j'$. That means,
the number of wave functions must be doubled after the passage of each avoided
crossing region. Therefore, treating the more realistic situations, the system
of coupled channel equations becomes much more complicated.
For simplicity, in our approximations we considered only a
superposition of $n$ diabatic wave functions, that means, the diabatic wave function
$i$ is forced to contribute to the amplitude of the diabatic wave function $j$
(which is not always equivalent to $j'$).
After some calculations, as detailed in Appendix \ref{appendixb}, a new set of TDHFB equations results:
\begin{equation}
i\hbar \dot{\rho}_{l(m)}=\kappa_{l(m)}\Delta_{m}^{*}-
\kappa_{l(m)}^{*}\Delta_{m},
\label{hfb21}
\end{equation}
\begin{equation}
i\hbar \dot{\kappa}_{l(m)}=\left(2\rho_{l(m)}-1\right)\Delta_{m}+
2\kappa_{l(m)}\left(\epsilon_{l}-\lambda_{m}\right),
\label{hfb22}
\end{equation}

\begin{equation}
i\hbar\dot{p}_{m}=\sum_{j\ne m}^{n}h_{mj}
\left(S_{mj}-
S_{jm},
\right)
\label{hfb23}
\end{equation}

\begin{equation}
\begin{array}{c}
i\hbar \dot{S}_{jm}=
S_{jm}\left\{-{1\over G}\left(\mid\Delta_{m}\mid^{2}-
\mid\Delta_{j}\mid^{2}\right)\right.\\
+(\epsilon_{m}(t)-\epsilon_{j}(t)-\lambda_{m}+\lambda_{j})\\
- {1\over 2}\sum_{k\ne m}\left(\Delta_{m}
\kappa_{k(m)}^{*}+\Delta_{m}^{*}
\kappa_{k(m)}\right)\left( {\rho_{k(m)}^{2}\over\mid\kappa_{k(m)}\mid^{2}}-1 \right)\\
+\left. {1\over 2}\sum_{k\ne j}\left(\Delta_{j}
\kappa_{k(j)}^{*}+\Delta_{j}^{*}
\kappa_{k(j)}\right)\left(  {\rho_{k(j)}^{2}\over\mid\kappa_{k(j)}\mid^{2}}-1       \right)    \right\}
\\
   +  \sum_{l\ne m,j}^{n}\left[h_{ml}(t)S_{jl}
    -  h_{jl}(t)S_{lm}\right]
+h_{mj}(t)(p_{j}-p_{m}).
\end{array}
\label{hfb24}
\end{equation}
The same notations as in the previous approach are used.

Two main differences arise between Eqs. (\ref{tbcs1})-(\ref{tbcs4}) 
and (\ref{hfb21})-(\ref{hfb24}) that are implicitly determined 
by the hypothesis assumed in their derivation. Firstly, in Eqs. (\ref{tbcs1})-(\ref{tbcs4}) 
the values of
$\dot{\rho}$ and $\dot{\kappa}$ are obtained through a weighted sum that runs over
unpaired states 
while in Eqs. (\ref{hfb21})-(\ref{hfb22}) these quantities belong to only one diabatic 
wave function.
 Secondly,  in Rel. (\ref{hfb24}), $\dot{S_{jm}}$
depends on all the densities $\rho$ and pairing moment components $\kappa$ of 
the implied two diabatic wave functions $j$ and $m$. As a consequence of these
differences, the number of differential 
equations increases
$n$ time in Eqs. (\ref{hfb21})-(\ref{hfb22}). Another consequence is that
2$\sum_{k\ne m}\rho_{k(m)}=N-1$ for each diabatic wave function $m$ in Eqs. (\ref{hfb21})-(\ref{hfb22})
while 2$\sum_{k}\rho_{k}=N+2\rho_{F}-1$ in Eqs. (\ref{tbcs1})-(\ref{tbcs4}) where $\rho_{F}$ 
denotes the single-particle density of the Fermi level in the initial ground-state configuration.
Finally, the chemical potential $\lambda$ has values associated
to the diabatic state under consideration in Eqs. 
(\ref{hfb21})-(\ref{hfb22}).

\section{ENERGY}

In this section, only the equations associated to the blocking level approach
are displayed. For the low lying level approach, the index $(m)$ 
must be dropped. 
The ground state energy $E_{0}$ of any deformation is obtained 
in the framework of the BCS
formalism by considering the Fermi level $\epsilon_{F}$ populated 
with the unpaired nucleon:
\begin{equation}
E_{0}=2\sum_{k\ne F}\rho_{k(F)}\epsilon_{k}+\epsilon_{F}-
G\mid\sum_{k\ne F} \kappa_{k(F)}\mid^{2}-G\sum_{k\ne F}\rho_{k(F)}^{2}.
\end{equation}
in the static, lower energy state.
For the same deformation,
the energy of an adiabatic state $m$ is obtained by considering the 
unpaired nucleon located on the diabatic state under consideration:
\begin{equation}
E_{m}=2\sum_{k\ne m}\rho_{k(m)}\epsilon_{k}+\epsilon_{m}-
G\mid\sum_{k\ne m} \kappa_{k(m)}\mid^{2}-G\sum_{k\ne m}\rho_{k(m)}^{2},
\end{equation}
where the solutions of the TDHFB equations are used.
In the frame of our model, the difference
\begin{equation}
\Delta E_{m}=E_{m}-E_{0},
\label{specialization}
\end{equation}
behaves as a specialization energy. So, as inferred
in Ref. \cite{wheeler} the quantity $\Delta E_{m}$ must increase the potential
barrier tunneled by the nuclear system. Different barriers are obtained
for each diabatic state under consideration. These appear as dynamic excitations
during the decaying process. Combining excitations with occupation probabilities
of diabatic states, we obtain 
\begin{equation}
{E}=\sum_{m}^{n}p_{m} E_{m},
\end{equation}
for the average energy and
\begin{equation}
{\Delta \bar E}=\sum_{m}^{n}p_{m}\Delta E_{m},
\label{average}
\end{equation}
for the averaged dissipated energy during the decay. As mentioned
in Ref. \cite{norem}, the collective kinetic energy is temporarily
stored as a conservative potential. This energy subsequently decays
partially to the dissipation.

The equations (\ref{tbcs1})-(\ref{tbcs4}) and (\ref{hfb21})-(\ref{hfb24})
involves only single-particle energies. They conserve the average 
number of particles because $2\sum_{k\ne m}\rho_{k(m)}=N-1$
for any $m$ (or
2$\sum_{k}\rho_{k}=N+2\rho_{F}-1$) 
and $\sum_{m}^{n}p_{m}=1$. The average energy can evolves in time as follows:
\begin{equation}
\begin{array}{c}
\dot{E}=\sum_{m}^{n}p_{m}
\left\{ 2\sum_{k\ne m}\rho_{k(m)}\dot\epsilon_{k}+\dot\epsilon_{m}\right.\\
-
\left.\dot{G}\mid\sum_{k\ne m} \kappa_{k(m)}\mid^{2}-\dot{G}
\sum_{k\ne m}\rho_{k(m)}^{2}
\right\}.
\end{array}
\end{equation}
For a stationary system, for which $\dot\epsilon$=0 and $\dot{G}$=0, the total 
energy is conserved, even if individual values of $p$, $\rho$  and $\kappa$ may still be varying
with time. 
In our treatment, the chemical potential has the values 
%$\lambda_{m}={E_{m}\over N}$.
$\lambda_{m}$ obtained from BCS equations for each energy levels workspace associated to
the diabatic wave function $m$.

\section{GENERALIZATION}

If the blocked levels are eliminated, the system (\ref{tbcs1})-(\ref{tbcs4}) reduces to: 
\begin{equation}
\begin{array}{c}
i\hbar\dot{\rho}_{l}=\kappa_{l}\Delta^{*}-\kappa_{l}^{*}\Delta,\\
i\hbar\dot{\kappa}_{l}=(2\rho_{l}-1)\Delta-
2\kappa_{l}\left[\epsilon_{l}(t)-\lambda(t)\right],
\end{array}
\label{bcsn}
\end{equation}	
the well known TDHFB equations \cite{koonin,blocki}. On another hand if the pairing 
is neglected, the third equation of the system (\ref{tbcs3}) can be written
\begin{equation}
i\hbar\left(\dot{c}_{m}c_{m}^{*}+\dot{c}_{m}^{*}c_{m}\right)=
\sum_{j\ne m}^{n}h_{mj}\left(c_{j}c_{m}^{*}+c_{j}^{*}c_{m}\right).
\end{equation}
Introducing explicitly the time dependence of the amplitudes $c_{m}$    
	\begin{equation}
c_{m}(t)=c_{0m}(t)\exp\left(-{i\over\hbar}\int_{0}^{t}\epsilon_{m}(\tau)d\tau\right),
\end{equation}
the next relation is obtained:
\begin{equation}
\begin{array}{c}
i\hbar(\dot{c}_{0m}c_{0m}^{*}+\dot{c}_{0m}^{*}c_{0m})\\
=\sum_{j\ne m}^{n}h_{jm}\left[c_{0j}c_{0m}^{*}\exp\left(-{i\over\hbar}\int_{0}^{t}
(\epsilon_{j}-\epsilon_{m})d\tau\right)\right.\\
-\left.c_{0j}^{*}c_{0m}\exp\left({i\over\hbar}\int_{
0}^{t}
(\epsilon_{j}-\epsilon_{m})d\tau\right)\right].
\end{array}
\end{equation}
The last relation is an equivalent form of the Landau-Zener Eq. (\ref{lzu}) 
obtained in the frame of the single-particle model. Furthermore, if the 
pairing interaction is neglected, $\rho$ and $\kappa$ can be either zero or one, and 
the fourth equation of the system (\ref{tbcs4}) reduces to
\begin{equation}
\begin {array}{c}
i\hbar\left(\dot{c}_{m}c_{j}^{*}+\dot{c}_{j}^{*}c_{m}\right)=
c_{j}^{*}c_{m}(\epsilon_{m}-\epsilon_{j})\\
+\sum_{l\ne m,j}^{n}
\left(h_{ml}c_{j}^{*}c_{l}-h_{jl}c_{l}^{*}c_{m}\right)+
h_{jm}\left(c_{j}^{*}c_{j}-c_{m}^{*}c_{m}\right).
\end{array}
\end{equation}	
After introducing the exponential dependence, the next relation emerge:
\begin{equation}
\begin{array}{c}
i\hbar\left(\dot{c}_{0m}c_{0j}^{*}+\dot{c}_{0j}^{*}c_{0m}\right)\exp\left(-{i\over\hbar}
\int(\epsilon_{m}-\epsilon_{j})d\tau\right)\\
=\sum_{l\ne m,j}^{n}\left[h_{ml}c_{0j}^{*}c_{0l}\exp\left
(-{i\over\hbar}\int(\epsilon_{l}-\epsilon_{j})
d\tau\right)\right.\\
-\left. h_{jl}c_{0l}^{*}c_{0m}\exp\left(-{i\over\hbar}\int(\epsilon_{m}-\epsilon_{l})
d\tau\right)\right]\\
+h_{mj}\left(c_{0j}^{*}c_{0j}-c_{0m}^{*}c_{0m}\right).
\end{array}
\end{equation}
That is another form of the Landau-Zener relation (\ref{lzu}). So, the Landau-Zener 
equation for single particle systems (without residual interactions) and the TDHFB 
equations for quasiparticles are two particular 
cases of the coupled channel equations (\ref{tbcs1})-(\ref{tbcs4}). 
So, this system represents a generalization of the 
TDHFB equations in the case of seniority one nuclear systems. Similar arguments are valid also
for the system (\ref{hfb21})-(\ref{hfb24}).

\section{RESULTS}

To solve the TDHFB equations, only the variations of the
single-particle energies $\epsilon_{k}$
are needed. The simplest way to obtain the
evolutions of single-particle energies is to consider
a time-dependent single particle potential in which the nucleons
move independently. As evidenced in Ref. \cite{koonin},
such a description is within the spirit of the more rigorous
Hartree-Fock approximation, which defines the potential 
self-consistently.

The $^{14}$C emission from $^{223}$Ra will be treated. The fragments issued in
this reaction are spherical while the parent is little deformed, allowing
a description in terms of a nuclear shape parametrization given by two
spheres smoothly joined within a third surface.

A fine structure in the $^{14}$ radioactivity of the $^{223}$Ra was
observed in 1989 \cite{brillard,hussonnois,hourany}. In the first experiment, the results
indicates that 15$\pm 3$ \% of $^{14}$C decays are transitions on the ground
state of the daughter, while 81$\pm 6$\% are transitions on the first excited 
state. In Ref. \cite{gupta}, using the M3Y potential, it was evidenced that the preformation
probability must be more favorable for the excited state than for the ground state
with a factor of 180. Such a value cannot be accounted from theoretical models \cite{deli1}
without taking into account dynamical ingredients. This is the main reason that 
the fine structure phenomenon was selected 
to validate our equations.

The deformation energy of the nuclear system is the sum between the
liquid drop energy and the shell effects, including pairing corrections.
The macroscopic energy is obtained in the framework of the 
Yukawa-plus-exponential model extended for binary systems with different
charge densities \cite{poen1}. The Strutinsky prescriptions \cite{brac1}
were computed on the basis of a new version of the superasymmetric
two-center shell model. This version solves a Woods-Saxon 
potential in terms of the two-center prescriptions as detailed 
in Appendix \ref{appendixc}.

Because the pairing equations diverges for an infinite number
of active levels, a limited number of levels are used
in the calculations: 31 levels above and 31 levels under the
the unpaired Fermi level in the initial ground state configuration
that is, $N-1$=62.
These levels are selected in terms of the spin projection $\Omega$
on the symmetry axis and kept as a single particle energies workspace.
A constant value of the pairing parameter $G=0.13$ MeV is used. 

\begin{figure}
\resizebox{0.50\textwidth}{!}{
  \includegraphics{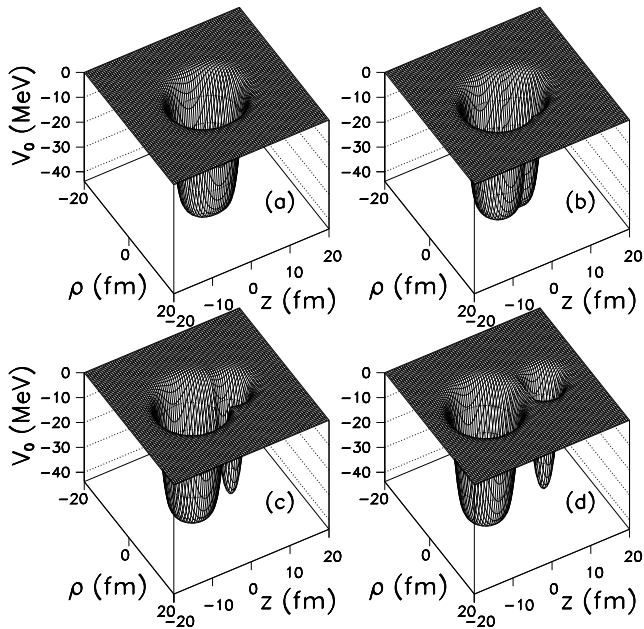}}
\caption{Mean field Woods-Saxon potential $V_{0}$ as function of the
cylindrical coordinates $\rho$ and $z$ for different values
of the elongation along the minimal action trajectory.
(a)Elongation  $R$=2 fm;
(b)$R$=5 fm; (c) $R$=10 fm; and (d) $R$=15 fm.}
\label{ws}
\end{figure}

The least action trajectory was obtained by generalizing in a three-dimensional
space the method
initiated in Ref. \cite{dnpmir} and used extensively to describe the fission 
processes \cite{fis1,fis2,fis3}. The inertia is computed within the
Werner-Wheeler method. The trajectory of the decaying system
is obtained simultaneously as function of three 
generalized coordinates, that is, the elongation
$R$ (the distance between the centers of the nascent fragments), the necking
parameter $C=S/R_{3}$ (the curvature of the intermediate surface)
and $R_{1}/R_{2}$ (the ratio between the radii of the heavy fragment $R_{1}$
and that of the light one $R_{2}$). These parameters are explained in
Appendix \ref{appendixc}. In Fig. \ref{potens}(a) the 
deformation energy $V$ of the nucleus
is plotted as function of the elongation $R$. Three excitations of the nuclear
systems that correspond to three adiabatic wave functions are also plotted with
dotted lines. These excitations are added to the deformation energies
obtained in the framework of the macroscopic-microscopic model
in order to calculate the penetrabilities as show below. 
In Fig. \ref{potens} (b) and (c), the
variations of the necking and mass-asymmetry generalized coordinates are displayed.
At $R\approx 10$ fm, a system formed by two spherical tangent nuclei is
obtained. The Woods-Saxon potential is presented in Fig. \ref{ws} for a
sequence of nuclear shapes along the least action path.

\begin{figure}
 \resizebox{0.50\textwidth}{!}{
   \includegraphics{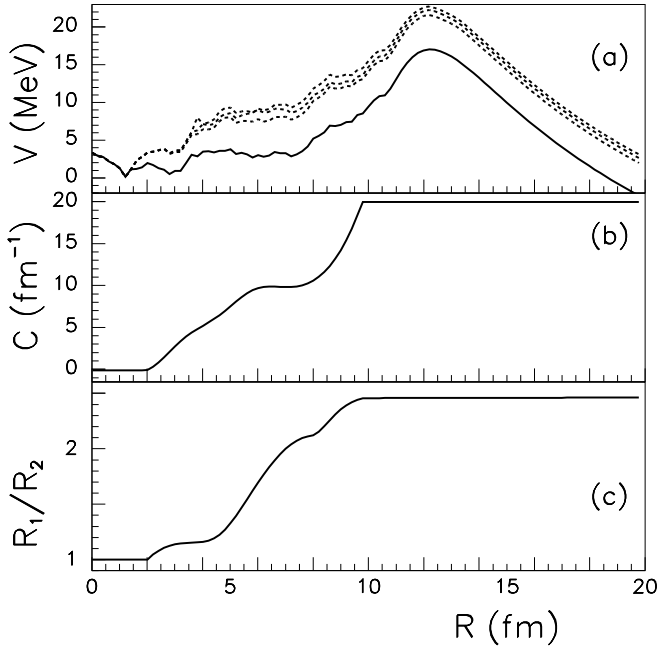}}
\caption{ (a) Deformation energy $V$ as function of the distance between the centers
of the nascent fragments $R$. The three excitations due to the diabatic levels
$\epsilon_{i}$, $i=1,3$ are also plotted with dotted lines. 
(b) Variation of the curvature of the
median surface and (c) of the mass-asymmetry parameter as function of $R$. }
\label{potens}
\end{figure}

\begin{figure}
\resizebox{0.50\textwidth}{!}{
  \includegraphics{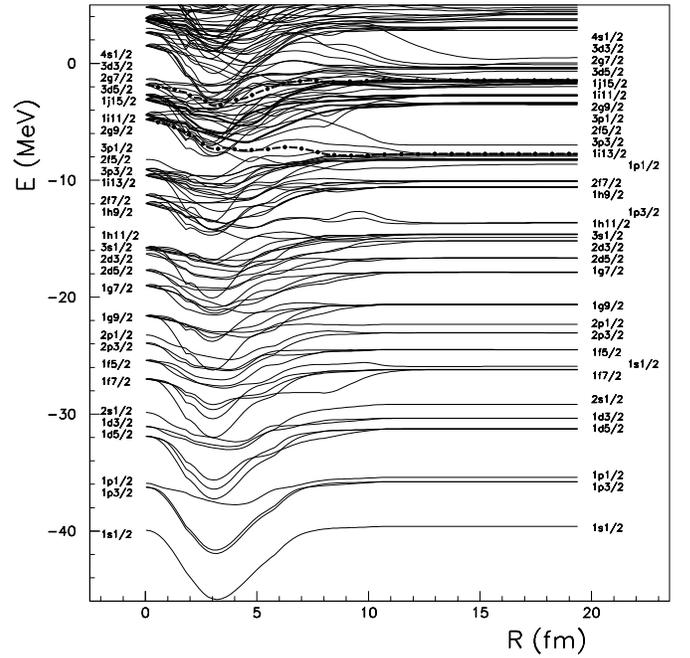}}
\caption{Neutron energy diagram along the minimal action path as
function of the distance between the centers of the fragments $R$.
The levels with spin projection $\Omega$ of interest are plotted with thick lines.
The levels are labeled within the spectroscopic factors. At the right the first
column corresponds to the daughter nucleus while the second one is related to the
$^{14}$C.}
\label{schema}
\end{figure}

\begin{figure}
\resizebox{0.50\textwidth}{!}{
  \includegraphics{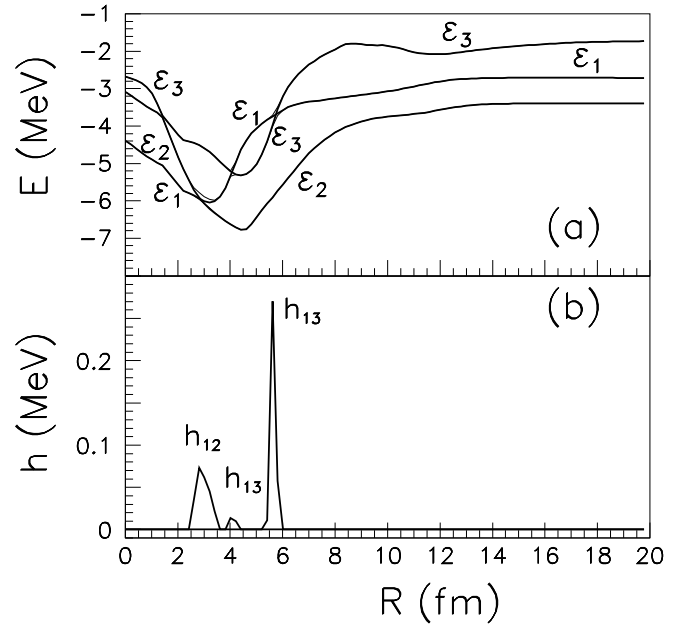}}
\caption{(a) Selected neutron energy levels that can be occupied by a single neutron
as function of the internuclear distance. Thick lines are the diabatic levels $\epsilon_{i}$,
$i$=1-3,
while thin lines are used for the adiabatic ones. (b) Interactions energies 
$h_{ij}$ in the avoided crossing regions.
}
\label{htt}
\end{figure}

\begin{figure}
\resizebox{0.50\textwidth}{!}{
  \includegraphics{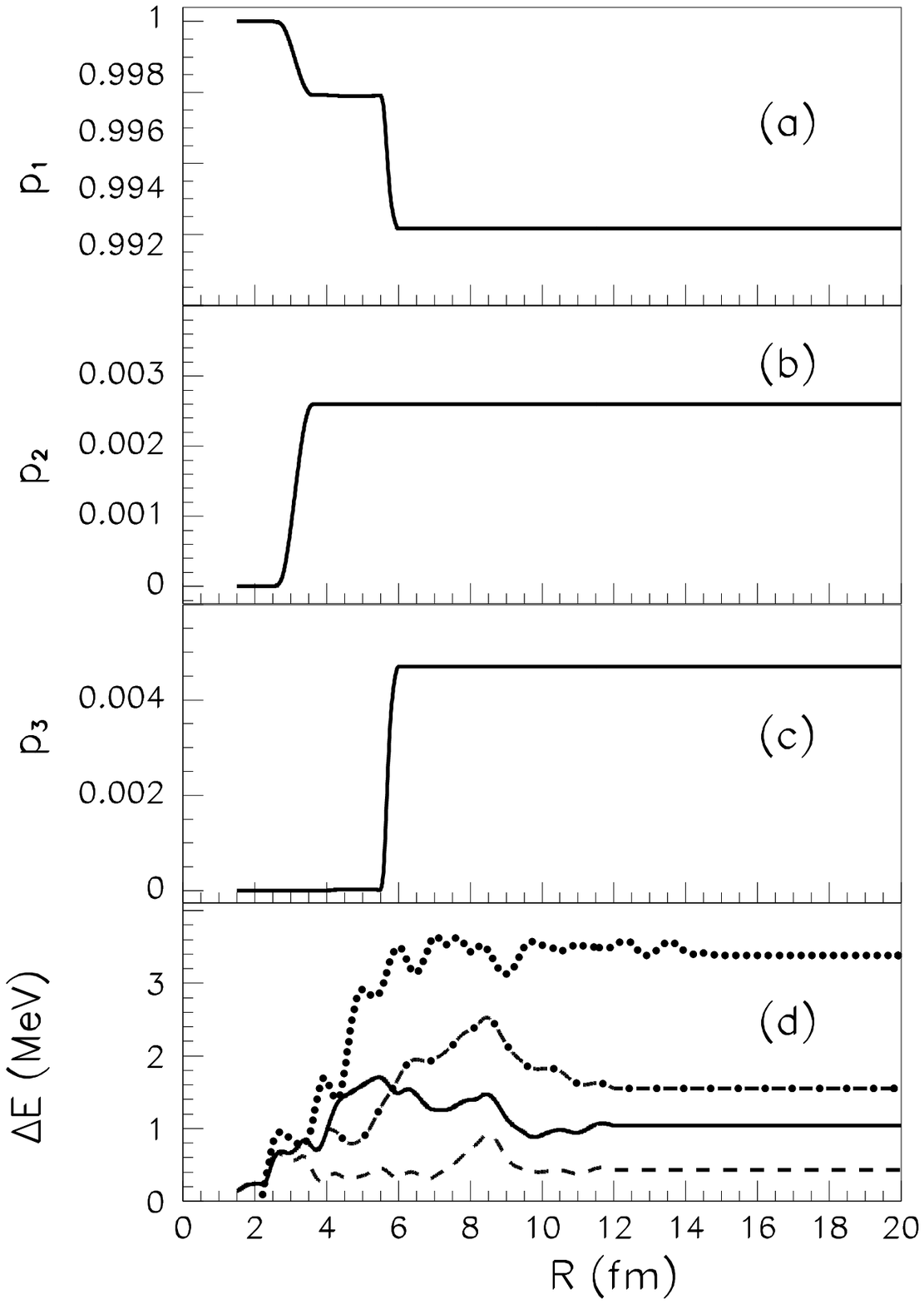}}
\caption{(a) Occupation probability $p_1$ for the diabatic 
level $\epsilon_{1}$ as function
of the internuclear distance $R$.
(b) Occupation probability $p_2$ of $\epsilon_2$. (c) Occupation probability 
$p_3$ of 
$\epsilon_3$. (c) The solid line corresponds to the dissipated energy 
$\Delta E_{1}$ for $\epsilon_{1}$, 
the dashed line is the dissipated 
energy $\Delta E_2$ for $\epsilon_2$, the dot-dashed line is the 
dissipation energy $\Delta E_3$ for $\epsilon_3$, and
the dotted line is the dissipation energy $\Delta E_p$ 
for the proton level scheme.
}
\label{probabili}
\end{figure}

The single-particle level schemes for neutrons and protons must be 
computed along the minimal action trajectory in order to solve the
time dependent pairing equations.
It is known that $^{223}$Ra has the spin ${3\over 2}$ emerging from 1$i_{11/2}$.
Adiabatically, the unpaired neutron reaches the $2g_{9/2}$ level of the daughter $^{209}$Pb.
As also evidenced in Refs. \cite{mirea14c,mireamod}, the fine structure in the
$^{14}$C radioactivity can be understood by an enhanced transition
probability of the unpaired neutron from the adiabatic level $\Omega=3/2$ emerging
from 1$i_{11/2}$ to the adiabatic level with the same spin projection $\Omega$
that emerges from $1j_{15/2}$, in terms of the Landau-Zener effect. The level
scheme of Fig. \ref{schema} shows that (adiabatically)
for $\Omega=3/2$ the $1i_{11/2}$ level reaches the
$2g_{9/2}$ daughter state, the $1j_{15/2}$ level arrives on the $1i_{11/2}$ one.
In this respect, the level scheme calculated within the Woods-Saxon model
is in qualitative agreement with that obtained within 
the modified oscillator model \cite{mirea14c}.
The $\Omega=3/2$ levels  subjected to avoided level crossings that can give rise to
Landau-Zener effect are plotted with thick lines in Fig. \ref{schema}.
Two adjacent levels with $\Omega=3/2$ are also plotted with point-dotted lines
to show that no other avoided level crossings are possible if the unpaired neutron
originates from $1i_{11/2}$. Our goal is to compute
the occupation probabilities of the 3 levels of interest at the end of the
disintegration process. For this purpose, Eqs. (\ref{hfb22}) - (\ref{hfb24}) are used.
Some features concerning the less rigorous low lying levels approach 
(\ref{tbcs1})-(\ref{tbcs2}) can be
found in Ref. \cite{mireamod}

In Fig. \ref{htt}, the three selected diabatic levels $\epsilon_{m}$ $(m=1,2,3)$ are
plotted together with the interaction energies $h_{ij}$ determined by using spline
interpolations around level crossings. Diabatically, the unpaired neutron,
initially located on the level $\epsilon_{1}$ that starts from the spherical orbital
$1i_{11/2}$ will arrive on the final state $1i_{11/2}$, that is the
first single particle excited state of the daughter, after the passage of three avoided
level crossing regions.

The initial conditions are determined by solving the BCS equations for the
3 possible seniority one wave functions at $R\approx$1.5 fm, where 
the first minimum of the deformation energy is located. The time dependent pairing
equations are integrated numerically using the Runge-Kutta method.
The occupation probabilities $p_{m}$ and the dissipated energies 
given by formula (\ref{specialization})
are determined along the minimal action path for a 
internuclear velocity ${\partial R\over\partial t}=1.4\times 10^{6}$ m/s. This value can be 
translated in a time required to penetrate the
barrier of about $1.4\times 10^{-20}$ s.

In Fig. \ref{probabili} the probability of occupations $p_{m}$ 
with an unpaired neutron
of the three diabatic levels
are presented as function of the internuclear distance. In the bottom panel,
the three dissipated energies $\Delta E_{m}$ and the energy dissipated in the
proton level scheme are also displayed. 

The branching ratio $r$ between the partial half-life for transitions
to the ground state of the
daughter and the partial half-live to the first excited state is given by
\begin{equation}
r={p_{1}\exp(-K_{1})\over p_{2}\exp(-K_{2})},
\end{equation} 
where the index corresponds to the diabatic level $\epsilon_{1}$ or $\epsilon_{2}$,
and 
\begin{equation}
K_{m}={2\over\hbar}\int_{R_{\rm gs}}^{R_m}\sqrt{\mu [V(R)+D_{m}(R)-V(R_{\rm gs})]}dR
\end{equation}
are the WKB integrals. Here $R_{\rm gs}$ is the ground state elongation, 
$R_{m}$ is the exit point from the barrier for the channel $m$,
$V(R)$ is the
macroscopic-microscopic energy, $D_{m}(R)=\Delta E_{m}(R)+\Delta E_{p}(R)$ is the dissipated energy,
$\Delta E_{m}$ being the specialization energy given
by Rel. (\ref{specialization}), $\Delta E_{p}$ denoting the dissipated energy
of the proton subsystem, and $\mu$ is the reduced mass. $\Delta E_p$
is calculated with Eqs. (\ref{bcsn}). The barriers
obtained for $V(R)+D_{m}(R)$ are plotted in Fig. \ref{potens} (a) with dotted lines.
The experimental values of $r$ range between 5.4 and 5.9. Our theoretical
value is $r$=5, which is in an excellent agreement with experimental data.

In conclusion, two approaches that generalize the 
Landau-Zener equations for seniority one
superfluid systems are presented. The new formalism
is valid for any kind of mean field approximations that include
a monopole pairing field.
The equations that describe
our approaches offer information
about the spectroscopic amplitudes and the dissipated energies
in different final channels. The new equations were
used to reproduce the qualitative and quantitative features
of the fine structure phenomenon in cluster decay. 
Up to now, this phenomenon was not described adequately in the frame
theories that do not include dynamical ingredients, as
evidenced in Ref. \cite{mirgu}. Within the time
dependent pairing equations, a good agreement
with experimental data was obtained. A new version of the superasymmetric 
two center shell model based on a Woods-Saxon potential was developed
and used in this context.  

\begin{acknowledgements}

This work was partially supported in the frame of the IDEI program
of the Romanian Ministry of Education and Research.
\end{acknowledgements}

\appendix
\section{}
\label{appendixa}

The TDHFB equations when the blocking level is neglected will be derived in this Appendix.
Following the same prescriptions as in Ref. \cite{blocki}, using the time dependent
Hamiltonian (\ref{ham1})
within corrections (\ref{hprim1}), and the trial wave functions (\ref{functii1}),
 the expected value of the Lagrange function is obtained:
\begin{equation}
\begin{array}{c}
<\varphi \mid H-i\hbar{\partial\over \partial t}+H'-\lambda N\mid \varphi >\\
=\sum_{m}^{n}\mid c_{m}\mid ^{2} \{2\sum_{k\ne m} \mid v_{k}\mid^{2}
(\epsilon_{k}-\lambda)+(\epsilon_{m}-\lambda)\\
-G\mid \sum_{k\ne m}u_{k}^{*}v_{k}\mid^{2}\}-\\
i\hbar\sum_{m}^{n}\mid c_{m}\mid^{2} [\sum_{k\ne m}{1\over 2}(v_{k}^{*}\dot{v}_{k}-
\dot{v}_{k}^{*}v_{k})]-i\hbar\sum_{m}^{n}c_{m}^{*}\dot{c}_{m}\\
+\sum_{m,j\ne m}^{n}h_{mj}c_{m}^{*}c_{j}
\end{array}
\label{expr0a}
\end{equation}
The following identities were used:
\begin{equation}
\begin{array}{c}
<\varphi\mid \sum_{k}\epsilon_{k}(a_{k}^{+}a_{k}+a_{\bar{k}}^{+}a_{\bar{k}})\mid\varphi>\\
=
\sum_{m}^{n}\mid c_{m}\mid^{2}[\epsilon_{m}+2\sum_{k\ne m}\epsilon_{k}\mid v_{k}\mid^{2}];
\end{array}
\end{equation}
\begin{equation}
\begin{array}{c}
<\varphi\mid \sum_{kl}a_{k}^{+}a_{\bar{k}}^{+}a_{l}a_{\bar{l}}\mid\varphi>\\
=
\sum_{m}^{n}\mid c_{m}\mid^{2}\left(\sum_{k\ne m}\mid v_{k}\mid^{4}+\sum_{l\ne m}u_{l}v^{*}_{l}
\sum_{k\ne m}u_{k}v_{k}\right);
\end{array}
\label{appr1}
\end{equation}
\begin{equation}
<\varphi\mid \alpha_{i}^{+}\alpha_{j}+\alpha_{j}^{+}\alpha_{i}\mid\varphi>=
c_{i}^{*}c_{j}+c_{j}^{*}c_{i};
\end{equation}
and
\begin{equation}
\begin{array}{c}
<\varphi\mid {\partial\over\partial t}\mid\varphi>\\
=\sum_{m}^{n}\left[c_{m}^{*}\dot{c}_{m}+\mid c_{m}\mid^{2}\sum_{k\ne m}
(u_{k}\dot{u}_{k}+v_{k}^{*}\dot{v}_{k})\right];
\end{array}
\end{equation}
because, as evidenced in Ref \cite{blocki},
\begin{equation}
\int(\phi_{k}^{*}\dot{\phi}_{k}+\dot{\phi}_{k}^{*}\phi_{k})d^{3}r=0,
\end{equation}
where $\mid\phi_{k}>=a_{k}^{+}\mid 0>$, due to the normalization.
The high order term $\mid v_{k}\mid^{4}$ of formula (\ref{appr1}) is neglected. 
The equality $\dot{u}_{k}=-(\dot{v}_{k}^{*}v_{k}+\dot{v}_{k}v_{k}^{*})/(2u_{k})$
is also used. 

To minimize the functional, the expression (\ref{expr0a}) is derived with respect 
to the independent variables $v_{l}^{*}$ and $v_{l}$. Two equations follow:
\begin{equation}
\begin{array}{c}
\sum_{m}^{n}\mid c_{m}\mid^{2}
\left\{2v_{l}^{*}(\epsilon_{l}-\lambda)-  
G[\sum_{k\ne m}\kappa_{k}\left(-{v_{l}^{*}v_{l}^{*}\over 2u_{l}}\right)\right.\\
\left.+\left(u_{l}-{\rho_{l}\over 2u_{l}}\right)\sum_{k\ne m}\kappa_{k}^{*}]
+i\hbar \dot{v}_{l}^{*}\right\}=0,
\label{cond01a}
\end{array}
\end{equation}

\begin{equation}
\begin{array}{c}
\sum_{m}^{n}\mid c_{m}\mid^{2}
\left\{2v_{l}(\epsilon_{l}-\lambda)-   
G[\sum_{k\ne m}\kappa_{k}^{*}\left(-{v_{l}v_{l}\over 2u_{l}}\right)\right.\\
\left.+\left(u_{l}-{\rho_{l}\over 2u_{l}}\right)\sum_{k\ne m}\kappa_{k}]
-i\hbar \dot{v}_{l}\right\}=0,
\label{cond02a}
\end{array}
\end{equation}
where the notations for densities $\rho=\mid v \mid^{2}$ and pairing moment components
$\kappa=uv$ are introduced.

The condition of conservation
of the number of particles
\begin{equation}
2\sum_{k}\mid v_{k}\mid^{2}=N+2\rho_{F}-1,
\end{equation}
was used so that
\begin{equation}
\sum_{k}(\dot{v}_{k}^{*}v_{k}+v_{k}^{*}\dot{v}_{k})=0,
\end{equation}
\begin{equation}
{\partial \over \partial v_{k}}(\dot{v}_{k}^{*}v_{k}+v_{k}^{*}\dot{v}_{k})=0,
\end{equation}
and
\begin{equation}
{1\over 2}{\partial\over\partial v_{l}}v_{k}^{*}\dot{v}_{l}=-\dot{v}_{l}^{*}.
\label{cosv}
\end{equation}

Multiplying Eqs. (\ref{cond01a}) and (\ref{cond02a}) with $v_{l}$ and
$v_{l}^{*}$, respectively,
and subtracting, the first TDHFB equation (\ref{tbcs1}) is obtained:
\begin{equation}
i\hbar \dot{\rho}_{l}={\sum_{m}^{n}\mid c_{m}\mid^{2}
\left\{\kappa_{l}\Delta_{m}^{*}-\kappa_{l}^{*}\Delta_{m}\right\}\over
\sum_{m}^{n}\mid c_{m}\mid^{2}},
\end{equation}
where $\sum_{m}\mid c_{m}\mid^{2}=1$ and $\Delta_{m}=G\sum_{k\ne m}\kappa_{k}$.

Another equation can be obtained:
\begin{equation}
\begin{array}{c}
\dot{\kappa}_{l}=-{v_{l}\over 2u_{l}}\dot{\rho}_{l}+u_{l}\dot{v}_{l}
=-{v_{l}\over 2u_{l}}{1\over i\hbar}\sum_{m}^{n}\mid c_{m}\mid^{2}\\
\times
\left\{\kappa_{l}\Delta_{m}^{*}-\kappa_{l}^{*}\Delta_{m}\right\}
+{u_{l}\over i\hbar}
\sum_{m}^{n}\mid c_{m}\mid^{2}\\
\times
\left[2v_{l}(\epsilon_{l}-\lambda)-\Delta^{*}_{m}
\left(-{v_{l}v_{l}\over 2u_{l}}\right)-\left(u_{l}-{\rho_{l}\over 2u_{l}}\right)
\Delta_{m}\right],
\end{array}
\end{equation}
so that the second TDHFB equation (\ref{tbcs2}) follows: 
\begin{equation}
i\hbar\dot\kappa_{l}=\sum_{m}^{n}\mid c_{m}\mid^{2}\left\{\left(2\rho_{l}-1\right)\Delta_{m}+
2\kappa_{l}\left(\epsilon_{l}-\lambda\right)\right\}.
\end{equation}
Using the property
\begin{equation}
\sum_{m}^{n}\mid c_{m}\mid^2=1;
\end{equation}
so that
\begin{equation}
\sum_{m}^{n}\dot{c}_{m}c_{m}^{*}=-\sum_{m}^{n}\dot{c}_{m}^{*}c_{m},
\end{equation}
the Eq. (\ref{expr0a}) is derived with respect $c_{m}$ and $c_{m}^{*}$ and set to zero. The 
next relations are obtained:
\begin{equation}
\begin{array}{c}
-i\hbar\dot{c}_{m}^{*}=c_{m}^{*}[2\sum_{k\ne m}\mid v_{k}\mid^{2}(\epsilon_{k}-\lambda)+
(\epsilon_{m}-\lambda)\\
-G\mid\sum_{k\ne m}u_{k}^{*}v_{k}\mid^{2}]\\
-i\hbar c^{*}_{m}[\sum_{k\ne m}{1\over 2}(v_{k}^{*}\dot{v}_{k}-\dot{v}_{k}^{*}v_{k})]
+\sum_{j\ne m}^{n}h_{mj}c_{j}^{*},
\end{array}
\label{eqr1}
\end{equation}

\begin{equation}
\begin{array}{c}
i\hbar\dot{c}_{m}=c_{m}[2\sum_{k\ne m}\mid v_{k}\mid^{2}(\epsilon_{k}-\lambda)+
(\epsilon_{m}-\lambda)\\
-G\mid\sum_{k\ne m}u_{k}^{*}v_{k}\mid^{2}]\\
-i\hbar c_{m}[\sum_{k\ne m}{1\over 2}(v_{k}^{*}\dot{v}_{k}-\dot{v}_{k}^{*}v_{k})]
+\sum_{j\ne m}^{n}h_{mj}c_{j},
\end{array}
\label{eqr2}
\end{equation}
Multiplying the relations (\ref{eqr1}) and (\ref{eqr2}) with $c_{m}$ and $c_{m}^{*}$ and
subtracting them the next relation follows
\begin{equation}
i\hbar(\dot{c}_{m}c_{m}^{*}+\dot{c}_{m}^{*}c_{m})=\sum_{j\ne m}^{n}h_{mj}
(c_{j}c_{m}^{*}-c_{j}^{*}c_{m}).
\label{eq88}
\end{equation}
It is a form of the third TDHFB equation (\ref{tbcs3}).
For a passage through only one avoided crossing region $(m,j)$, 
only two amplitudes,
$c_{m}$ and $c_{j}$ can change. On another hand, from the condition
of conservation it can be obtained:
\begin{equation}
\dot{c}_{m}c_{m}^{*}+c_{m}\dot{c}_{m}^{*}=
-(\dot{c}_{j}c_{j}^{*}+c_{j}\dot{c}_{j}^{*})
\end{equation}
This condition is fulfilled by the above Eq. (\ref{eq88}), so the equation
conserves the norm.
Changing indexes, multiplying with amplitudes and subtracting relations 
(\ref{eqr1}) and (\ref{eqr2}) the next equation follows:
\begin{equation}
\begin{array}{c}
i\hbar(\dot{c}_{m}c_{j}^{*}+\dot{c}_{j}^{*}c_{m})=
c_{m}c_{j}^{*}\left\{-{1\over G}\right(\mid\Delta_{m}\mid^{2}-\mid\Delta_{j}\mid^{2}\left)+
\right.\\
2\mid v_{j}\mid^{2}(\epsilon_{j}-\lambda)+
(\epsilon_{m}-\lambda)-\Delta_{m}\\
-\left. 2\mid v_{m}\mid^{2}(\epsilon_{m}-\lambda)-
(\epsilon_{j}-\lambda)-\Delta_{j}\right\}\\
+c_{m}c_{j}^{*}{i\hbar\over 2}(v_{m}^{*}\dot{v}_{m}^{*}-\dot{v}_{m}^{*}v_{m}-
v_{j}^{*}\dot{v}_{j}+\dot{v}_{j}^{*}v_{j})
\end{array}
\end{equation}
From Eqs. (\ref{cond01a}) and (\ref{cond02a}), the expressions in the last
parenthesis that involves $\dot{v}$ and $\dot{v}^{*}$ can be obtained: 
\begin{equation}
\begin{array}{c}
{i\hbar\over 2}(v_{l}^{*}\dot{v}_{l}-\dot{v}_{l}^{*}v_{l})
=2\rho_{l}(\epsilon_{l}-\lambda)\\
+{\sum_{m}^{n}p_{m}\Delta_{m}\over 2}\left({\rho_{l}^{2}\over \kappa_{l}}-\kappa_{l}^{*}\right)+
{\sum_{m}^{n}p_{m}\Delta_{m}^{*}\over 2}\left({\rho_{l}^{2}\over \kappa_{l}^{*}}-\kappa_{l}\right),
\end{array}
\end{equation}
Using the
notations (\ref{notatii}) and rearranging the terms
the fourth TDHF equation (\ref{tbcs4}) is obtained.

\section{}
\label{appendixb}

The TDHFB equations when the blocking effect is taken into consideration are derived
in this Appendix. Using the corrections (\ref{cort}), and the trial wave functions 
(\ref{wf2}), the expected value of the Lagrange function is:
\begin{equation}
\begin{array}{c}
<\varphi \mid H-i\hbar{\partial\over \partial t}+H'-\lambda N\mid \varphi >\\
=\sum_{m}^{n}\mid c_{m}\mid ^{2} \{2\sum_{k\ne m} \mid v_{k(m)}\mid^{2}
(\epsilon_{k}-\lambda_{m})\\
+(\epsilon_{m}-\lambda_{m})-G\mid \sum_{k\ne m}u_{k(m)}v_{k(m)}\mid^{2}\}\\
-i\hbar\sum_{m}^{n}\mid c_{m}\mid^{2} [\sum_{k\ne m}{1\over 2}(v_{k(m)}^{*}\dot{v}_{k(m)}
-\dot{v}_{k(m)}^{*}v_{k(m)})]\\
-i\hbar\sum_{m}^{n}c_{m}^{*}\dot{c}_{m}
+\sum_{m,j\ne m}^{n}h_{mj}c_{m}^{*}c_{j}
\end{array}
\label{expr0}
\end{equation}
In order to minimize the functional, the expression (\ref{expr0}) is derived with respect 
to the independent variables $v_{l(m)}^{*}$ and $v_{l(m)}$ by taking into account 
the subsidiary condition
(\ref{cosv}). The next relations follows:
\begin{equation}
\begin{array}{c}
\sum_{m}^{n}\mid c_{m}\mid^{2}
\left\{2v_{l(m)}^{*}(\epsilon_{l}-\lambda_{m})  \right. \\
-G\left[\sum_{k\ne m}\kappa_{k(m)}
\left(-{v_{l(m)}^{*}v_{l(m)}^{*}\over 2u_{l(m)}}\right)\right.\\
+\left.\left.\left(u_{l(m)}-{\rho_{l(m)}\over 2u_{l(m)}}
\right)\sum_{k\ne m}\kappa_{k(m)}^{*}\right]
+i\hbar \dot{v}_{l(m)}^{*}\right\}=0,
\end{array}
\label{cond0}
\end{equation}

\begin{equation}
\begin{array}{c}
\sum_{m}^{n}\mid c_{m}\mid^{2}
\left\{2v_{l(m)}(\epsilon_{l}-\lambda_{m})  \right. \\
-G\left[\sum_{k\ne m}\kappa_{k(m)}^{*}
\left(-{v_{l(m)}v_{l(m)}\over 2u_{l(m)}}\right)\right.\\
+\left.\left.\left(u_{l(m)}-{\rho_{l(m)}\over 2u_{l(m)}}
\right)\sum_{k\ne m}\kappa_{k(m)}\right]
-i\hbar \dot{v}_{l(m)}\right\}=0.
\end{array}
\label{cond00}
\end{equation}
This system can be solved by considering
that the expression in the curly bracket is zero for each value
of $m$. Following a similar way as in Appendix 1, the first two TDHFB 
equations associated to an unpaired nucleon in the state
$m$ emerge:
\begin{equation}
i\hbar \dot{\rho}_{l(m)}=\kappa_{l(m)}\Delta_{m}^{*}-
\kappa_{l(m)}^{*}\Delta_{m},
\end{equation}
\begin{equation}
i\hbar \dot{\kappa}_{l(m)}=\left(2\rho_{l(m)}-1\right)\Delta_{m}+
2\kappa_{l(m)}\left(\epsilon_{l}-\lambda_{m}\right).
\end{equation}
To obtain the probability that an unpaired nucleon is located on
a state $m$, the expression
(\ref{expr0}) must be derived with respect $c_{m}$ and $c_{m}^{*}$.
Two equations follow: 
\begin{equation}
\begin{array}{c}
-i\hbar \dot{c}_{m}^{*}=
c_{m}^{*}\left[2\sum_{k\ne m}\mid v_{k(m)}\mid^{2}(\epsilon_{k}-\lambda_{m})+
\epsilon_{m}-\lambda_{m}\right.\\
-G\mid\sum_{k\ne m}u_{k(m)}v_{k(m)}\mid^{2}\\
-\left. i\hbar c_{m}^{*}\sum_{k\ne m}{1\over 2}
(v_{k(m)}^{*}\dot{v}_{k(m)}-\dot{v}_{k(m)}^{*}v_{k(m)})\right]\\
+\sum_{j\ne m}^{n}h_{mj}c_{j}^{*}=0,
\end{array}
 \label{fro19}
\end{equation}
\begin{equation}
\begin{array}{c}
i\hbar \dot{c}_{m}=
c_{m}\left[2\sum_{k\ne m}\mid v_{k(m)}\mid^{2}(\epsilon_{k}-\lambda_{m})+
\epsilon_{m}-\lambda_{m}\right.\\
-G\mid\sum_{k\ne m}u_{k(m)}v_{k(m)}\mid^{2}\\
-\left. i\hbar c_{m}\sum_{k\ne m}{1\over 2}
(v_{k(m)}^{*}\dot{v}_{k(m)}-\dot{v}_{k(m)}^{*}v_{k(m)})\right]\\
+\sum_{j\ne m}^{n}h_{mj}c_{j}=0.
\end{array}
 \label{fro20}
\end{equation}
Multiplying with complex conjugates and subtracting, the next relation 
\begin{equation}
\begin{array}{c}
i\hbar (\dot{c}_{m}c_{m}^{*}+\dot{c}_{m}^{*}c_{m})\\
=\sum_{j\ne m}^{n}h_{mj}\left(c_{j}c_{m}^{*}-c_{j}^{*}c_{m}\right),
\end{array}
\end{equation}
is obtained.
Using the notations (\ref{notatii}) the Eq. (\ref{hfb23}) follows. 

From relations (\ref{fro19}) and (\ref{fro20}), another relation can be deduced
\begin{equation}
\begin{array}{c}
i\hbar(\dot{c}_{j}^{*}c_{m}+\dot{c}_{m}c_{j}^{*})\\
=c_{m}c_{j}^{*}\left[-{1\over G}\left(\mid\Delta_{m}\mid^{2}-\mid\Delta_{j}\mid^{2}\right)
+(\epsilon_{m}-\epsilon_{j}-\lambda_{m}+\lambda_{j})\right.\\
\left. +2\sum_{k\ne m}\rho_{k(m)}(\epsilon_{k}-\lambda_{m})-
2\sum_{k\ne j}\rho_{k(j)}(\epsilon_{k}-\lambda_{j})\right]\\
-{i\hbar\over 2} c_{m}\dot{c}_{j}^{*}
\left[\sum_{k\ne m}(v_{k(m)}^{*}\dot{v}_{k(m)}-\dot{v}_{k(m)}^{*}v_{k(m)})\right.\\
-\left.
\sum_{k\ne j}(v_{k(j)}^{*}\dot{v}_{k(j)}-\dot{v}_{k(j)}^{*}v_{k(j)})\right]\\
+\sum_{l\ne m}^{n}h_{lm}c_{l}c_{j}^{*}-
\sum_{l\ne j}^{n}h_{lj}c_{l}^{*}c_{m}.
\end{array}
\label{mmmm}
\end{equation}
The derivatives $\dot{v}$ and $\dot{v}^{*}$ appear in the previous expression.
In order to evaluate quantities where these derivatives intervene, 
Eqs. (\ref{cond0}) and (\ref{cond00}) are used. The next relation follows.
\begin{equation}
\begin{array}{c}
{i\hbar\over 2}(v_{l(m)}^{*}\dot{v}_{l(m)}-\dot{v}_{l(m)}^{*}v_{l(m)})\\
={v_{l(m)}^{*}\over2}\{(2v_{l(m)}(\epsilon_{l}-\lambda_{m})\\
-G[\sum_{k\ne m}\kappa_{k(m)}^{*}(-{v_{l(m)}^{*}v_{l(m)}v_{l(m)}\over 2v_{l(m)}^{*}u_{l(m)}}\\
+(u_{l(m)}-{\rho_{l(m)}\over 2u_{l(m)}})\sum_{k\ne m}\kappa_{k(m)}]\}\\
+{v_{l(m)}\over2}\{(2v_{l(m)}^{*}(\epsilon_{l}-\lambda_{m})\\
-G[\sum_{k\ne m}\kappa_{k(m)}(-{v_{l(m)}v_{l(m)}^{*}v_{l(m)}^{*}\over 2v_{l(m)}u_{l(m)}}\\
+(u_{l(m)}-{\rho_{l(m)}\over 2u_{l(m)}})\sum_{k\ne m}\kappa_{k(m)}^{*}]\}\\
=\rho_{l(m)}(\epsilon_{l}-\lambda_{m})+{\Delta_{m}^{*}\over 2}
{\rho_{l(m)}^{2}\over 2\kappa_{l(m)}^{*}}-
(\kappa_{l(m)}^{*}-{\rho_{l(m)}^{2}\over 2\kappa_{l(m)}}){\Delta_{m}\over2}\\
+\rho_{l(m)}(\epsilon_{l}-\lambda_{m})+{\Delta_{m}\over 2}{\rho_{l(m)}^{2}\over
2\kappa_{l(m)}}-
(\kappa_{l(m)}-{\rho_{l(m)}^{2}\over 2\kappa_{l(m)}^{*}}){\Delta_{m}^{*}\over2}\\
=2\rho_{l(m)}(\epsilon_{l}-\lambda_{m})\\
+{\Delta_{m}^{*}\over 2}\left({\rho_{l(m)}^{2}\over \kappa_{l(m)}^{*}}-\kappa_{l(m)}\right)+
{\Delta_{m}\over 2}\left({\rho_{l(m)}^{2}\over \kappa_{l(m)}}-\kappa_{l(m)}^{*}\right),
\end{array}
\end{equation}
so that relation (\ref{mmmm}) becomes:
\begin{equation}
\begin{array}{c}
i\hbar \dot{S}_{jm}=c_{m}c_{j}^{*}\left\{-{1\over G}\left(\mid\Delta_{m}\mid^{2}-
\mid\Delta_{j}\mid^{2}\right)\right.\\
+(\epsilon_{m}-\epsilon_{j}-\lambda_{m}+\lambda_{j})\\     %%%%%%
- {1\over 2}\sum_{k\ne m}\left[\Delta_{m}\left({\rho_{k(m)}^{2}\over\kappa_{k(m)}}-
\kappa_{k(m)}^{*}\right)+\Delta_{m}^{*}\left({\rho_{k(m)}^{2}\over\kappa_{k(m)}^{*}}-
\kappa_{k(m)}\right)\right]\\                                %%%%%%%%%%%%%%%
+\left. {1\over 2}\sum_{k\ne j}\left[\Delta_{j}\left({\rho_{k(j)}^{2}\over\kappa_{k(j)}}-
\kappa_{k(j)}^{*}\right)+\Delta_{j}^{*}\left({\rho_{k(j)}^{2}\over\kappa_{k(j)}^{*}}-
\kappa_{k(j)}\right)\right]\right\}\\                       %%%%%%%%%%%%%%
+\sum_{k\ne j,m}^{n}h_{mk}c_{k}c_{j}^{*}
-\sum_{k\ne j,m}^{n}h_{kj}c_{k}^{*}c_{m}\\
+h_{mj}c_{j}c_{j}^{*}
-h_{jm}c_{m}^{*}c_{m}.
\end{array}
\end{equation}
After some rearrangements of terms, and using notations
(\ref{notatii}) the Eq. (\ref{hfb24}) is obtained.

\section{}
\label{appendixc}

\begin{figure}[bth]
\resizebox{0.50\textwidth}{!}{
  \includegraphics{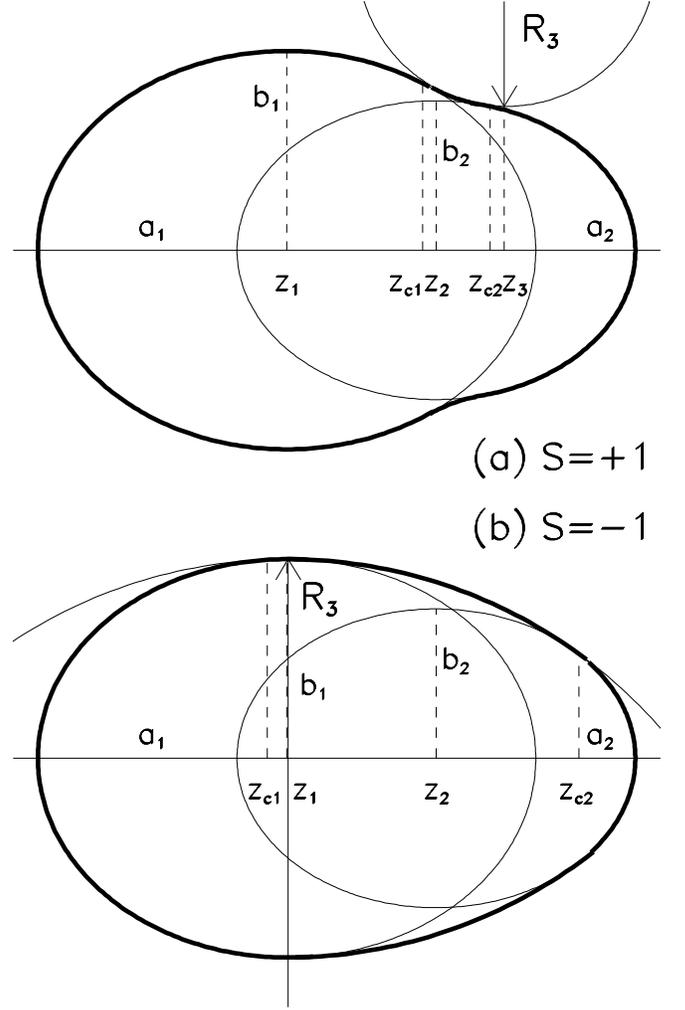}}
%\centerline{\includegraphics[width=8cm]{parametr.eps}}
\caption{Nuclear shape parametrization.
Two intersected ellipsoids of different eccentricities are smoothly
joined with a third surface. Two cases can be obtained: (a) the curvature
of the circle of radius $R_{3}$ is positive ($s$=1) and
(b) the curvature of $R_{3}$ is negative ($s$=-1). The elongation is given
by the distance between the centers of the ellipsoids $R=z_{2}-z_{1}$}
\label{parametr}
\end{figure}

A two-center shell model with a Woods-Saxon potential was developed recently \cite{mirws}.
An  axial symmetric nuclear shape parametrization is used to determine the mean field potential.
This nuclear shape parametrization is given by two ellipsoids (of different semi-axis and 
eccentricities)
smoothly joined with a third surface given
by the rotation of a circle around the axis of symmetry as displayed in
Fig. \ref{parametr}. The parametrization is characterized by 5 degrees of freedom
that can be associated, for example,  to 
the elongation ($R=z_{2}-z_{1}$), to the necking ($C=S/R_{3}$), to the mass asymmetry
($\eta=a_{1}/a_{2}$), to the deformations of the two fragments ($b_{i}/a_{i}$, $i=1,2$).
Treating the $^{14}$C emission, the deformations of the two fragments can be
neglected and the mass asymmetry parameter is considered as $\eta=R_{1}/R_{2}$.
The mean field potential is defined in the frame of the Woods-Saxon model:
\begin{equation}
V_{0}(\rho,z)=-{V_{c}\over 1+\exp\left[{\Delta(\rho,z)\over a}\right]}
\end{equation}
where $\Delta(\rho,z)$ represents the distance 
between a point $(\rho,z)$
and the nuclear surface. This distance is measured only along
the normal direction on the surface and it is negative if the point $(\rho,z)$ is located
in the
interior of the nucleus. $V_{c}$ is the depth of the
potential while $a$ is the diffuseness parameter. In our work, the
depth is $V_{c}=V_{0c}[1\pm \kappa (N_{0}-Z_{0})/N_{0}+Z_{0})]$ with plus sign for
protons and minus sign for neutrons, $V_{0c}$= 51 MeV, $a$=0.67 fm,
$\kappa$=0.67.  Here $A_{0}$, $N_{0}$ and $Z_{0}$ represent the
mass number, the neutron number and the charge number of the parent, respectively.
This parametrization, referred as the Blomqvist-Walhlborn one 
in Ref. \cite{scwiok}, is adopted 
because it provides the same
radius constant $r_{0}$ for the mean field and the pairing field. 
That ensures a consistency of the shapes of the two fields at 
hyperdeformations, i.e., two tangent ellipsoids.

In Fig. \ref{ws}, the mean field potential $V_{0}$ is plotted as function of
cylindrical coordinates 
$\rho$ and $z$ for four nuclear shape
configurations obtained along the minimal action path.

The spin-orbit coupling is assumed of the form
\begin{equation}
V_{ls}=-2\lambda\left({1\over 2mc}\right)^{2}(\nabla V_{0}\times \vec{p})\vec{s}
\end{equation}
where $\lambda$=35 is a dimensionless coupling constant, $m$ is the
nucleon mass while $c$ denotes the speed of the light.
  The spherical components of the operator
\begin{equation}
L=\nabla V\times p
\end{equation}
in cylindrical coordinates are
\begin{equation}
L^{\pm}=\mp \hbar e^{\pm i\varphi}
\left({\partial V_{0}\over \partial\rho}{\partial\over \partial z}
-{\partial V_{0}\over \partial z}{\partial\over \partial\rho}
\pm i {\partial V_{0}\over\partial z}{1\over \rho}{\partial\over\partial\varphi}
\right)
\end{equation}
\begin{equation}
L_{z}=i\hbar{\partial V_{0}\over\partial\rho}{1\over\rho}{\partial
\over\partial\varphi}
\end{equation}
so that
\begin{equation}
Ls={1\over 2}(L^{+}s^{-}+L^{-}s^{+})+L_{z}s_{z}
\end{equation}

The next step is to obtain the solutions of the 
Schr\"{o}dinger equation
\begin{equation}
\begin{array}{c}
\left[-{\hbar^{2}\over 2m}\Delta
+V_{0}(\rho,z)+V_{ls}(\rho,z)+V_{C}(\rho,z)\right]\Psi(\rho,z,\varphi)\\
=
E\Psi(\rho,z,\varphi)
\end{array}
\end{equation}
For protons, a Coulomb term $V_{C}$ is added as in Ref. \cite{scwiok}.
No analytical solutions can be found for such
potentials. A suitable eigenvector basis able to
diagonalize the Woods-Saxon potential can be obtained with
the double center harmonic oscillator model.

A complete analytical eigenvector basis can be only obtained
for the semi-symmetric two-center oscillator. This
potential corresponds to a shape parametrization given
by two ellipsoids that possess the same semi-axis perpendicular
on the axis of symmetry. The potential
is
\begin{equation}
V_{o}(\rho,z)=\left\{\begin{array}{lc}
{1\over2}m\omega_{z1}^{2}(z-c_{1})^{2}+{1\over2}m\omega_{\rho}^{2},& z<0,\\
{1\over2}m\omega_{z2}^{2}(z-c_{2})^{2}+{1\over2}m\omega_{\rho}^{2},& z\ge0,
\end{array}\right.
\end{equation}
where $\omega$ denotes the stiffness of the potential along
different directions as follows, $\omega_{z1}=\omega_{0}{R_{0}\over a_{1}}$,
$\omega_{z2}=\omega_{0}{R_{0}\over a_{2}}$, $\omega_{\rho}=\omega_{0}{R_{0}\over b_{1}}$,
$\omega_{0}=41A_{0}^{-1/3}$, $R_{0}=r_{0}A_{0}^{1/3}$, in order to 
ensure a constant value of the potential on the surface. The origin
on the $z$-axis is considered the location of the plane of 
intersection between the two ellipsoids.

An analytic system of eigenvectors can be
obtained for $V_{0}$ by solving the   Schr\"{o}dinger equation:
\begin{equation}
\left[-{\hbar^{2}\over 2m_{0}}\Delta+V_{o}(\rho,z)\right]\Psi(\rho,z,\varphi)=E\Psi(\rho,z,\varphi)
\label{eqsp2}
\end{equation}
The analytic solution of Eq. (\ref{eqsp2}) is obtained using the ansatz
\begin{equation}
\Psi(\rho,z,\varphi)=Z(z)R(\rho)\Phi(\varphi)
\label{eivb}
\end{equation}
with
\begin{equation}
\Phi_{m}(\varphi)={1\over\sqrt{2\pi}}\exp(im\varphi)
\end{equation}
\begin{equation}
R_{nm}(\rho)=\sqrt{2n!\over (n+m)!}\alpha_{\rho}\exp\left(-{\alpha_{\rho}^{2}\rho^{2}\over 2}\right)
(\alpha_{\rho}\rho)^{m}L_{n}^{m}(\alpha_{\rho}^{2}\rho^{2})
\end{equation}
\begin{equation}
Z_{\nu}(z)=\left\{ \begin{array}{l}
C_{\nu_{1}}\exp\left(-{\alpha_{z1}^{2}(z-c_{1})^{2}\over 2}\right){\bf{H}}_{\nu_{1}}[-\alpha_{z1}(z+c_{1})],\\~~~~~~~~z<0,\\
C_{\nu_{2}}\exp\left(-{\alpha_{z2}^{2}(z-c_{2})^{2}\over 2}\right){\bf{H}}_{\nu_{2}}[\alpha_{z2}(z-c_{2})],\\~~~~~~~z\ge 0,
\end{array}\right.
\label{basis}
\end{equation}
where $L_{n}^{m}(x)$ is the Laguerre polynomial, ${\bf{H}}_{\nu}(\zeta)$ is the Hermite function,
$\alpha_{i}=(m_{0}\omega_{i}/\hbar)^{1/2}$ ($i=z1,z2,\rho$) are length parameters, 
and $C_{\nu_{i}}$ denote the normalization constants.
The quantum numbers $n$ and $m$ are integers while the quantum number $\nu$ along the $z$-axis is real
and has different values for the intervals $(-\infty,0]$ and $[0,\infty)$.
Imposing conditions for the continuity of the wave function and its derivative, together
with those for the stationary energy and orthonormality, the values of $\nu_{1}$, $\nu_{2}$, $C_{\nu_{1}}$ and
$C_{\nu_{2}}$ are obtained.
Details concerning these solutions and expressions for the normalization constants 
are found in Refs. \cite{mireatc,mireatc2}.
For reflection-symmetric shapes, the solutions along the $z$-axis are also characterized 
by the parity as a good quantum number. 
The basis (\ref{basis}) for the two-center
oscillators can be used for a various ranges of models 
more of less phenomenological \cite{maruhn,geng}.
On another hand, they are different ways
to obtain the single-particle energies for a two-center Woods-Saxon potential. Other
recipes are given in Ref. \cite{torres} where the potentials are expanded in terms
of harmonic oscillators functions.

\end{document}